\documentclass[conference]{IEEEtran}
\IEEEoverridecommandlockouts
\usepackage{amssymb}
\usepackage{cite}
\usepackage{graphicx}
\usepackage{array}
\usepackage{multirow}
\usepackage{amsmath}
\usepackage{stfloats}
\usepackage{graphicx}
\usepackage{subfigure}
\usepackage{tabularx}
\usepackage{epsfig,epsf}
\usepackage{setspace}
\usepackage{bm}
\usepackage{textcomp}
\usepackage{algorithmic}
\usepackage{algorithm}
\usepackage{subfigure}
\usepackage{caption}
\usepackage{graphicx}
\usepackage{setspace}
\usepackage{url}
\usepackage{verbatim}
\usepackage{graphicx}
\usepackage{cite}
\usepackage{amssymb}
\usepackage{color}
\usepackage{xpatch}
\definecolor{shadecolor}{RGB}{0,0,255}
\definecolor{blue}{RGB}{0,0,255}

\newtheorem{lemma}{Lemma}

\makeatletter

\ExplSyntaxOn
\cs_new:Npn \bibColoredItems #1#2
{
	\clist_map_inline:nn {#2} { \cs_new:cpn {bib@colored@##1} {#1} } 
}
\ExplSyntaxOff

\newcommand\bib@setcolor[1]{%
	\ifcsname bib@colored@#1\endcsname
	\expanded{\noexpand\color{\csname bib@colored@#1\endcsname}}%
	\else
	\normalcolor
	\fi
}

\xpatchcmd\@bibitem
{\item}
{\bib@setcolor{#1}\item}
{}{\fail}

\xpatchcmd\@lbibitem
{\item}
{\bib@setcolor{#2}\item}
{}{\fail}
\makeatother

\begin{document}
	
\title{RIS-assisted Atomic MIMO Receiver}
 
\author{Qihao Peng, Jiuyu Liu, Qu Luo, ~\IEEEmembership{Member,~IEEE}, Yi Ma, ~\IEEEmembership{Senior Member,~IEEE},  \\ Pei Xiao, ~\IEEEmembership{Senior Member,~IEEE}, Maged Elkashlan, ~\IEEEmembership{Senior Member,~IEEE}, George K. Karagiannidis,  ~\IEEEmembership{Fellow,~IEEE}.
   
		\thanks{Q. Peng, J. Liu, Q, Luo, Y. Ma, and P. Xiao, are affiliated with 5G and 6G Innovation Centre, Institute for Communication Systems (ICS) of the University of Surrey, Guildford, GU2 7XH, UK. (e-mail: \{q.peng, jiuyu.liu, q.u.luo, y.ma, p.xiao\}@surrey.ac.uk). (\emph{Corresponding Author: Jiuyu Liu and Qu Luo})} \\
          \thanks{M. Elkashlan is with the School of Electrical Engineering and Computer Science of Queen Mary University of London. (e-mail: m.elkashlan@qmul.ac.uk).} \\
        \thanks{George K. Karagiannidis is with the Department of Electrical and Computer Engineering, Aristotle University of Thessaloniki, Greece. (email: geokarag@auth.gr). }}

	
	\maketitle

\begin{abstract}
	In this paper, we propose a novel and low-complexity atomic multiple-input multiple-output (MIMO) receiver architecture assisted by a reconfigurable intelligent surface (RIS). By introducing RIS and utilizing pulse amplitude modulation (PAM), the phase of the transmitted signal is effectively aligned with that of the local oscillator (LO), thereby mitigating phase ambiguity and substantially reducing both signal detection complexity and overall receiver complexity. To tackle the resulting non-convex optimization problem, we reformulate it into a tractable form by minimizing the Frobenius norm of an equivalent matrix, which is efficiently solved using an Adam-based gradient descent algorithm. Building upon the optimized equivalent signals, we develop a low-complexity signal detection algorithm and evaluate its performance. Simulation results demonstrate that the proposed RIS-assisted atomic MIMO receiver significantly enhances detection performance compared to the conventional atomic MIMO receivers, achieving a 4 dB gain for 4-PAM compared to the existing benchmarks. More importantly, these results underscore the proposed architecture’s potential for scalable and high-performance atomic MIMO communication systems.
\end{abstract}	
	

\section{Introduction}
Atomic multiple-input multiple-output (MIMO) receivers have emerged as a promising paradigm for next-generation wireless communication systems, leveraging the exceptional electromagnetic field sensitivity of Rydberg atoms \cite{sedlacek2012microwave,liu2025sa}. By exploiting electromagnetically induced transparency (EIT) and Autler-Townes (AT) splitting effects \cite{finkelstein2023practical}, atomic sensors convert radio-frequency (RF)-induced spectral variations into optical signals, thereby enabling high sensitivity, wide bandwidth, and excellent linearity \cite{gong2025rydberg}. As a result, atomic MIMO receivers hold great potential to complement conventional receivers in future communication and sensing systems \cite{gong2024rydberg}.

Building upon these remarkable physical properties, extensive studies have explored atomic MIMO receivers from both experimental and theoretical perspectives \cite{meyer2018digital,anderson2020atomic,yuan2023rydberg,cui2024mimo,cui2025towards}. On the experimental side, pioneering studies have successfully demonstrated atom-based receivers for amplitude modulation (AM) and frequency modulation (FM) \cite{meyer2018digital,anderson2020atomic}. Then, the AM-based reception was investigated for millimeter-wave wireless communication in \cite{yuan2023rydberg}. However, most existing prototypes are limited to point-to-point communication. To tackle this issue from the theoretical side, Cui \emph{et al.} proposed iterative signal detection algorithms \cite{cui2025towards} and precoding schemes \cite{cui2024mimo}, which is not applicable for high system load as multi-user reception remains challenging due to nonlinear distortions introduced by photodetectors (PDs). As a result, achieving scalable, low-complexity, and high-performance multi-user signal detection in atomic MIMO systems is still an open problem.

To address these challenges, we propose a novel and low-complexity reconfigurable intelligent surface (RIS)-assisted atomic MIMO receiver architecture. Specifically, we employ pulse amplitude modulation (PAM) and optimize the phase shifts of RIS to align the phase of the received signals with the local oscillator (LO), thereby mitigating phase ambiguity and simplifying signal detection and receiver architecture. Furthermore, we design an efficient optimization framework to support scalable multi-user communication. Extensive simulation results validate the effectiveness of the proposed architecture by significant performance gains over conventional atomic MIMO receivers, underscoring the potential of RIS-assisted atomic MIMO systems for future wireless networks.

The remaining sections are organized as follows. The system model is presented in Section II. The passive phase optimization and signal detection algorithm of the RIS-aided atomic receiver are devised in Section III. In Section IV, extensive numerical results are presented. Finally, our conclusions are drawn in Section V.

\section{System Model and Problem Statement}

\begin{figure}
	\centering
	\includegraphics[width=3.2in]{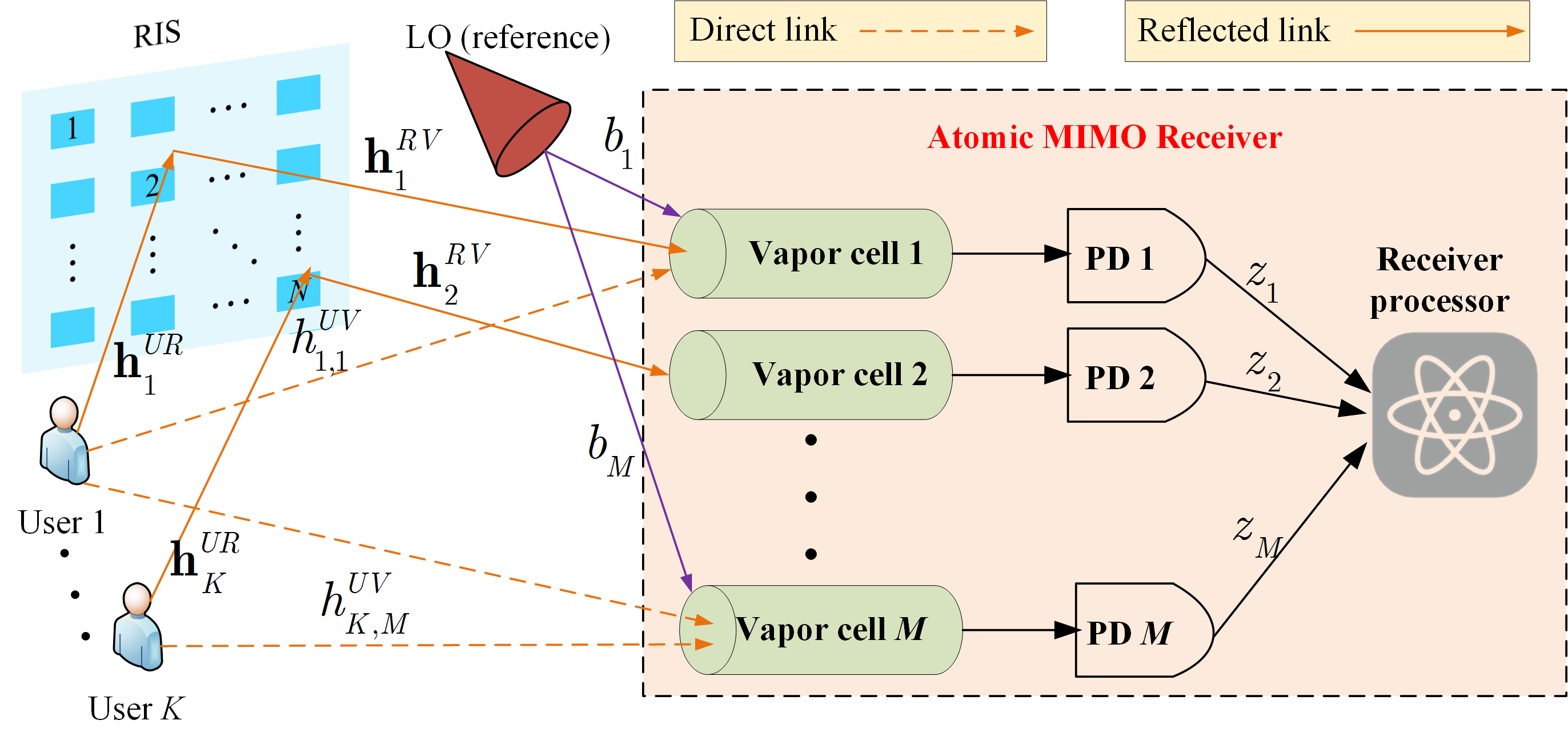}
	\caption{RIS-assisted atomic MIMO architecture, where the \(M\) atomic cells receive the signal from \(K\) users with the help of \(N\) RIS elements.}
	\label{system}
    \vspace{-0.8cm}
\end{figure}

\subsection{System Model}
In this section, we briefly present the system model of an atomic MIMO receiver, and a comprehensive overview of Rydberg atomic receivers can be found in \cite{cui2025towards}. As depicted in Fig.~\ref{system}, we propose a novel receiver, where \(K\) single-antenna users communicate with the atomic MIMO receiver through a reconfigurable intelligent surface (RIS) comprising \(N\) elements. The receiver is equipped with \(M\) vapor cells, which act as atomic antennas for measuring electromagnetic waves. Furthermore, each vapor cell is filled with Rydberg atoms and paired with a dedicated PD, enabling optical readout of the measured signals. The outputs of all PDs are subsequently aggregated and processed by a digital processor to perform signal detection.

The channel between the \(k\)-th user and RIS is denoted as \(\mathbf{h}^{\text{UR}}_k \in \mathbb{C}^{N \times 1}\), \(k \in \{1,\cdots, K\}\), and each element of \(\mathbf{h}^{\text{UR}}_k\) follows a complex Gaussian distribution \(\mathcal{CN}(0,1)\). Then, the channel matrix between \(K\) users and RIS \(\mathbf{H}^{\text{UR}} \in \mathbb{C}^{N \times K}\) is denoted as
\begin{equation}
    \small 
    \label{channelUR}
    \mathbf{H}^{\text{UR}} = [\mathbf{h}^{\text{UR}}_1,\cdots,\mathbf{h}^{\text{UR}}_K] .
\end{equation}
The channel between the vapor cells and the RIS is denoted as \(\mathbf{H}^{\text{UV}} \in \mathbb{C}^{M\times K}\), and its \((m,k)\)-th entry can be expressed as \cite{cui2025towards,cui2024mimo}
\begin{equation}
    \label{ChannelRV}
    \small
    [\mathbf{H}^{\text{UV}}]_{m,k} = \sum\nolimits_{l=1}^{L}\frac{1}{\hbar }\boldsymbol{\mu}^T_{\text{RF}}\boldsymbol{\epsilon}_{m,k,l}\rho_{m,k,l}e^{j\varphi_{m,k,l}},
\end{equation}
where \(\boldsymbol{\mu}_{\text{RF}} \in \mathbb{R}^{3 \times 1}\) represents the electric dipole moment of Rydberg atoms, \(L\) is the number of paths, and \(\hbar\) is the reduced Planck constant.  \(\boldsymbol{\epsilon}_{m,k,l} \in \mathbb{R}^{3 \times 1}\), \(\rho_{m,k,l}\), and \(\varphi_{m,k,l}\)  respectively denote the polarization direction, path loss, and phase shift of the radio wave propagating from the \(k\)-th user to the \(m\)-th atomic antenna via the \(l\)-th path. Similarly, by using the equation (\ref{ChannelRV}), the channel between the RIS and the atomic receiver can be denoted as \(\mathbf{H}^{\text{RV}}\in \mathbb{C}^{M \times N}\). By denoting the RIS's phase shift as \(\boldsymbol{\Phi} = \text{diag}\{e^{j\theta_1},\cdots,e^{j\theta_N}\}\), the effective channel matrix is \(\mathbf{H}^{\text{eq}} = \mathbf{H}^{\text{RV}}\mathbf{\Phi}\mathbf{H}^{\text{UR}}+\mathbf{H}^{\text{UV}}\). Finally, the received signals after PDs is given by \cite{cui2025towards,cui2024mimo,liu2025sa}
\begin{equation}
    \label{RXsignal}
    \mathbf{z} = |\mathbf{H}^{\text{eq}}\mathbf{s} + \mathbf{b} + \mathbf{n}|,
\end{equation}
where \(\mathbf{s} \in \mathbb{C}^{K \times 1}\) is the transmitted information, \(\mathbf{n}\) represents the noise following the complex Gaussian distribution of \(\mathcal{CN}(\mathbf{0},\sigma^2\mathbf{I}_M)\) owing to the law-of-large-number  \cite{cui2024mimo,cui2025towards} , and \(|x|\) implies the magnitude of \(x\). \(\mathbf{b} \in \mathbb{C}^{M \times 1}\) is the local oscillator (LO), and the \(m\)-th element \(b_m = \frac{s_b}{\hbar} \boldsymbol{\mu}^H_{\text{eg}}\boldsymbol{\epsilon}_{b,m}\sqrt{P_b}\rho_{b,m}e^{j\varphi_{b,m}}\), where \(s_b\) is the known reference symbol, \(P_b\) is the power of LO and \(\boldsymbol{\mu}_{\text{eg}}\) is the electric dipole moment. Furthermore, \(\boldsymbol{\epsilon}_{b,m}\), \(\rho_{b,m}\), and \(\varphi_{b,m}\) imply the polarization direction, path loss, and phase shift of LO, respectively.

\subsection{Problem Statement}
As can be seen from (\ref{RXsignal}), the received signal can only be detected by magnitude, while the wireless channel and information are both complex. To address this issue, we adopt the RIS for dynamically adjusting the passive phase shift to transform the received signal into a complex channel aligned with the LO, which can be expressed as
\begin{equation}
    \label{alignphase}
    \angle (\mathbf{H}^{\text{RV}}\mathbf{\Phi}\mathbf{H}^{\text{UR}}+\mathbf{H}^{\text{UV}})\mathbf{s} = \angle \mathbf{b},
\end{equation}
where \(\angle \mathbf{x} \) represents the angle of  \(\mathbf{x}\). Then, by extracting the phase of \(\mathbf{H}^{\text{eq}}\mathbf{s}\), we have 
\begin{equation}
\label{HEQ}
    \mathbf{H}^{\text{eq}}\mathbf{s}= \mathbf{H}^{\text{opt}}\mathbf{s} \circ e^{j\angle \mathbf{b}},
\end{equation}
 where \(\mathbf{H}^{\text{opt}} \in \mathbb{R}^{M \times K} \) represents the real channel after phase shift optimization and \(\circ\)  denotes the Hadamard product. By substituting (\ref{HEQ}) into (\ref{RXsignal}), the received signal can be rewritten as
\begin{equation}
    \mathbf{z} = |\mathbf{H}^{\text{opt}}\mathbf{s} \circ e^{j\angle \mathbf{b}} + \mathbf{b} + \mathbf{n}|.
\end{equation}
By introducing the RIS, it is easy to find the initial phase and operate signal detection compared to methods in \cite{cui2025towards}, thereby reducing the complexity of the atomic MIMO receiver.

Based on the above discussions, the problem can be formulated as
\begin{subequations}
\label{optimization}
\begin{align}
\small
\mathop {\min }\limits_{\{\theta_1,\cdots,\theta_N\}} \quad & \left\|\Im\left\{(\mathbf{H}^{\text{RV}}\mathbf{\Phi}\mathbf{H}^{\text{UR}}+\mathbf{H}^{\text{UV}})\mathbf{s} \circ e^{-j\angle \mathbf{b}}\right\}\right\|^2_2 \notag\\
{\rm{s}}{\rm{.t}}{\rm{.}}\;\;\;\; & \theta_n \in [0, 2\pi), \forall n \in \{1, \cdots, N\},
\end{align}
\end{subequations}
where \(\Im(\mathbf{x})\) and \(\left\| \mathbf{x}\right\|_2\) denote the imaginary part and norm of \(\mathbf{x}\). However, it is challenging to adjust the phase shift aligned with the LO due to the unknown transmitted information of \(\mathbf{s}\).


\section{Proposed RIS-aided Atomic MIMO Receiver}
In this section, we optimize the phase shift of the RIS and design a low-complexity signal detection algorithm for the proposed atomic MIMO receiver.
\subsection{Passive Phase Shift Design}
To solve Problem (\ref{optimization}), we assume that each user utilizes PAM signals, i.e., \(\mathbf{s} \in \mathbb{R}^{K \times 1}\) with \(\mathbb{E}\{\mathbf{ss^T}\} = \mathbf{I}_K\). Then, we simplify the objective function in (\ref{optimization}) by using the following lemma.

\begin{lemma}
	\label{lemma1}
        With the given vectors, \(\mathbf{s} \in \mathbb{R}^{K \times 1}\) and \(\mathbf{b} \in \mathbb{C}^{M \times 1}\), and complex matrices, inclulding \(\mathbf{A} \in \mathbb{C}^{M \times N}\), \(\mathbf{B} \in \mathbb{C}^{N \times K}\), \(\mathbf{C} \in \mathbb{C}^{M \times K}\), and \(\mathbf{\Phi} \in \mathbb{C}^{N \times N}\), we can solve the following problem to obtain the solution for minimizing \(\left\|\Im\left\{(\mathbf{A}\mathbf{\Phi}\mathbf{B}+\mathbf{C})\mathbf{s} \circ e^{-j\angle \mathbf{b}}\right\}\right\|^2_2\), which can be expressed as
        \begin{equation}
        \begin{split}
         \mathop {\min }\limits_{\{\theta_1,\cdots,\theta_N\}} \left\|\Im\left\{(\mathbf{A}\mathbf{\Phi}\mathbf{B}+\mathbf{C})\right\}\right\|^2_F,
        \end{split}    
        \end{equation}
	where \(\left\| \mathbf{X} \right\|_F \) denotes the Frobenius norm of \(\mathbf{X}\).
	
	{\emph{Proof}}: See Appendix A. $\hfill\blacksquare$
\end{lemma}

Therefore, by using Lemma \ref{lemma1}, the problem of aligning the phase of the received signal with that of the reference signal can be equivalently reformulated as transforming \(\mathbf{H}^{\text{eq}}\) into a real-value channel, which can be formulated as
\begin{subequations}
\label{reoptimization}
\begin{align}
\small
\mathop {\min }\limits_{\{\theta_1,\cdots,\theta_N\}} \quad &\left\|\Im\left\{(\mathbf{H}^{\text{RV}}\mathbf{\Phi}\mathbf{H}^{\text{UR}}+\mathbf{H}^{\text{UV}})\right\}\right\|^2_F \notag\\
{\rm{s}}{\rm{.t}}{\rm{.}}\;\;\;\; & \theta_n \in [0, 2\pi), \forall n \in \{1, \cdots, N\},
\end{align}
\end{subequations}

To transform Problem (\ref{optimization}) to a tractable form, we rewrite the objective function as
\begin{equation}
    \label{rechannel}
    J(\boldsymbol{\theta}) = \left\|\Im\left\{\mathbf{H}^{\text{UV}} + \sum\limits_{n=1}^N e^{j\theta_n} \mathbf{H}^{\text{RV}}(:,n)\mathbf{H}^{\text{UR}}(n,:)\right\} \right\|_F^2, 
\end{equation}
where \(\mathbf{H}(:,n)\) and \(\mathbf{H}(n,:)\) denote the \(n\)-th column and the \(n\)-th row of \(\mathbf{H}\). Then, by defining \(\mathbf{V}_n\) as \(\mathbf{H}^{\text{RV}}(:,n)\mathbf{H}^{\text{UR}}(n,:)\), we have 
\begin{subequations}
\label{reoptimization}
\begin{align}
    \mathop {\min }\limits_{\{\theta_1,\cdots,\theta_N\}} \quad &   \sum\limits_{m=1}^{M}\sum\limits_{k=1}^K \Big\{\Im([\mathbf{H}^{\text{UV}}]_{m,k})+\sum\limits_{n=1}^N \Big(\cos{\theta_n}\times \notag\\
    &\Im([\mathbf{V}_n]_{m,k})+\sin{\theta_n}\Re([\mathbf{V}_n]_{m,k}) \Big)\Big\}^2 \notag\\
    {\rm{s}}{\rm{.t}}{\rm{.}}\;\;\;\; & \theta_n \in [0, 2\pi), \forall n \in \{1, \cdots, N\},
    \end{align}
\end{subequations}
where \(\Re(\mathbf{X})\) means the real part of \(\mathbf{X}\) and \([\mathbf{H}]_{m,k}\) denotes the \((m,k)\)-th entry of \(\mathbf{H}\). By calculating the first-order derivative of the objective function, we have 
\begin{equation}
    \label{firstorder}
    \begin{split}
            \frac{\partial  J(\boldsymbol{\theta})}{\partial \theta_n} &= 2 \sum\limits_{m=1}^{M}\sum\limits_{k=1}^K Q_{m,k}\Big[\sin{\theta_n}\Im([\mathbf{V}_n]_{m,k})\\
            &-\cos{\theta_n}\Re([\mathbf{V}_n]_{m,k})\Big], \\
    \end{split}
\end{equation}
\begin{equation}
\begin{split}
         Q_{m,k} &= \Im([\mathbf{H}^{\text{UV}}]_{m,k})+\sum\limits_{n=1}^N \Big(\cos{\theta_n}\times \\
    &\Im([\mathbf{V}_n]_{m,k})+\sin{\theta_n}\Re([\mathbf{V}_n]_{m,k}) \Big).
\end{split}
\end{equation}
Based on the first-order derivative, the phase shift can be readily obtained by using the gradient descent algorithm \cite{peng2024RIS}, which is detailed in Algorithm \ref{GDA_algorithm}.

\begin{algorithm}[t]
	\caption{Gradient Descent Algorithm for Solving Problem (\ref{reoptimization})}
	\begin{algorithmic}[1]
		\label{GDA_algorithm}
		\STATE Initilize \(n = 1\), \(N_{\max} = 100\), \(\epsilon = 10^{-5}\), \(\eta = 0.05\), \(\beta_1 = 0.9 \), and \(\beta_2 = 0.999\);
		\STATE Initialize the phase shifts \(\boldsymbol{\theta}_{(n)}\) randomly, \(\mathbf{m}_{(0)} = \mathbf{0}\), \(\mathbf{v}_{(0)} = \mathbf{0}\);
		\WHILE {\(n < N_{\max}\)} 
		\STATE Calculate the gradient vector \(\mathbf{g}(\boldsymbol{\theta}_{(n)}) = [\frac{\partial  J(\boldsymbol{\theta})}{\partial \theta_1},\cdots,\frac{\partial  J(\boldsymbol{\theta})}{\partial \theta_N}]^T \Big|_{\boldsymbol{\theta} = \boldsymbol{\theta}_{(n)}}\);
		\STATE \(\mathbf{m}_{(n)} = \beta_1 \mathbf{m}_{(n-1)} + (1-\beta_1)\mathbf{g}(\boldsymbol{\theta}_{(n)})\);
		\STATE \(\mathbf{v}_{(n)} = \beta_2\mathbf{v}_{(n-1)} + (1-\beta_2)\mathbf{g}(\boldsymbol{\theta}_{(n)}) \circ\mathbf{g}(\boldsymbol{\theta}_{(n)})\) ;
		\STATE \(\mathbf{\hat m}_{(n)} = \frac{\mathbf{m}_{(n)}}{1-(\beta_1)^n}\);
            \STATE \(\mathbf{\hat v}_{(n)} = \frac{\mathbf{v}_{(n)}}{1-(\beta_2)^n}\);
            \STATE \(\boldsymbol{\theta}_{(n+1)} = \boldsymbol{\theta}_{(n)} -\eta \frac{\mathbf{\hat m}_{(n)}}{\sqrt{\mathbf{\hat v}_{(n)}} + \epsilon}\)
		\STATE Update $n = n + 1$;
		\ENDWHILE
	\end{algorithmic}
\end{algorithm}
\vspace{-0.5cm}

\subsection{Signal Detection}
Based on the above discussions, the received signal \(\mathbf{z}\) can be given by
\begin{equation}
    \label{z1}
    \mathbf{z} = |(\mathbf{H}^{\text{opt}} \mathbf{s} + |\mathbf{b}|) \circ e^{j \angle\mathbf{b}} + \mathbf{n}|.
\end{equation}
Since the LO is near the vapor cell, the magnitude of the LO is larger than that of noise, i.e., \(|\mathbf{b}| \gg |\mathbf{n}|\), and thus we have
\begin{equation}
    \mathbf{z}  = |\mathbf{b}| + \mathbf{H}^{\text{opt}} \mathbf{s} + \mathbf{\hat n},
\end{equation}
where \(\mathbf{\hat n}\) is the residual noise. Based on these discussions, the detected information \( \mathbf{\hat s}\) can be expressed as
\begin{equation}
    \label{hats}
    \mathbf{\hat s} = ((\mathbf{H}^{\text{eq}})^H\mathbf{H}^{\text{eq}})^{-1} (\mathbf{H}^{\text{eq}})^H(\mathbf{z} \circ e^{j\angle\mathbf{b}}-\mathbf{b}).
\end{equation}

Compared to the methods proposed in \cite{cui2025towards}, our RIS-aided atomic MIMO receiver enables direct signal detection without requiring initial phase estimation or iterative signal detection, thereby significantly reducing the complexity of both the detection process and the receiver architecture \footnote{Although the RIS phase requires iterative optimization, this optimization is performed on a channel-based timescale, which operates at a much lower frequency than the transmitted signal. As a result, the processing time and computational complexity at the receiver are substantially reduced.}.

\subsection{Complexity Analysis}
For the convergence of Algorithm \ref{GDA_algorithm}, it has been proven that the gradient descent algorithm based on the moment can converge to a locally optimal solution \cite{kingma2014adam}, which is omitted for brevity. The complexity of our proposed receiver mainly depends on the number of iterations of passive phase shift optimization. Specifically, based on \cite{kingma2014adam}, each iteration for operating the gradient descent algorithm is \(\mathcal{O}(3N)\). Therefore, the total complexity of our proposed atomic MIMO receiver is \(\mathcal{O}(3NN_{\max})\), where \(N_{\max}\) is the maximum number of iterations.

\section{Simulation Results}
In this section, we present simulation results to validate the effectiveness of our proposed RIS-aided atomic MIMO receiver and evaluate the proposed signal detection algorithm with other benchmarks.

\subsection{Parameter Settings}
Unless otherwise specified, the simulation setup for the atomic MIMO receiver is configured as follows. The number of vapor cells is set to \(M = 36\), and the number of passive elements in the RIS is \(N = 150\). For modulation schemes, 4-PAM, 8-PAM, and 16-PAM are adopted in our proposed RIS-aided atomic MIMO system. For atomic setups, by utilizing the Rydberg calculator \cite{robertson2021arc}, the transition dipole moment \(\boldsymbol{\mu}_{\text{eg}}\) and electric dipole moment \(\boldsymbol{\mu}_{\text{RF}}\) can be obtained. Then, the polarization direction, i.e., \(\boldsymbol{\epsilon}_{b,m}\) and \(\boldsymbol{\epsilon}_{mkl}\), are randomly generated from unit circles perpendicular to their incident angles. Furthermore, the channel coefficients are the same as Table I given in \cite{cui2025towards}.

\subsection{Convergence}
\begin{figure}
	\centering
	\includegraphics[width=2.5in]{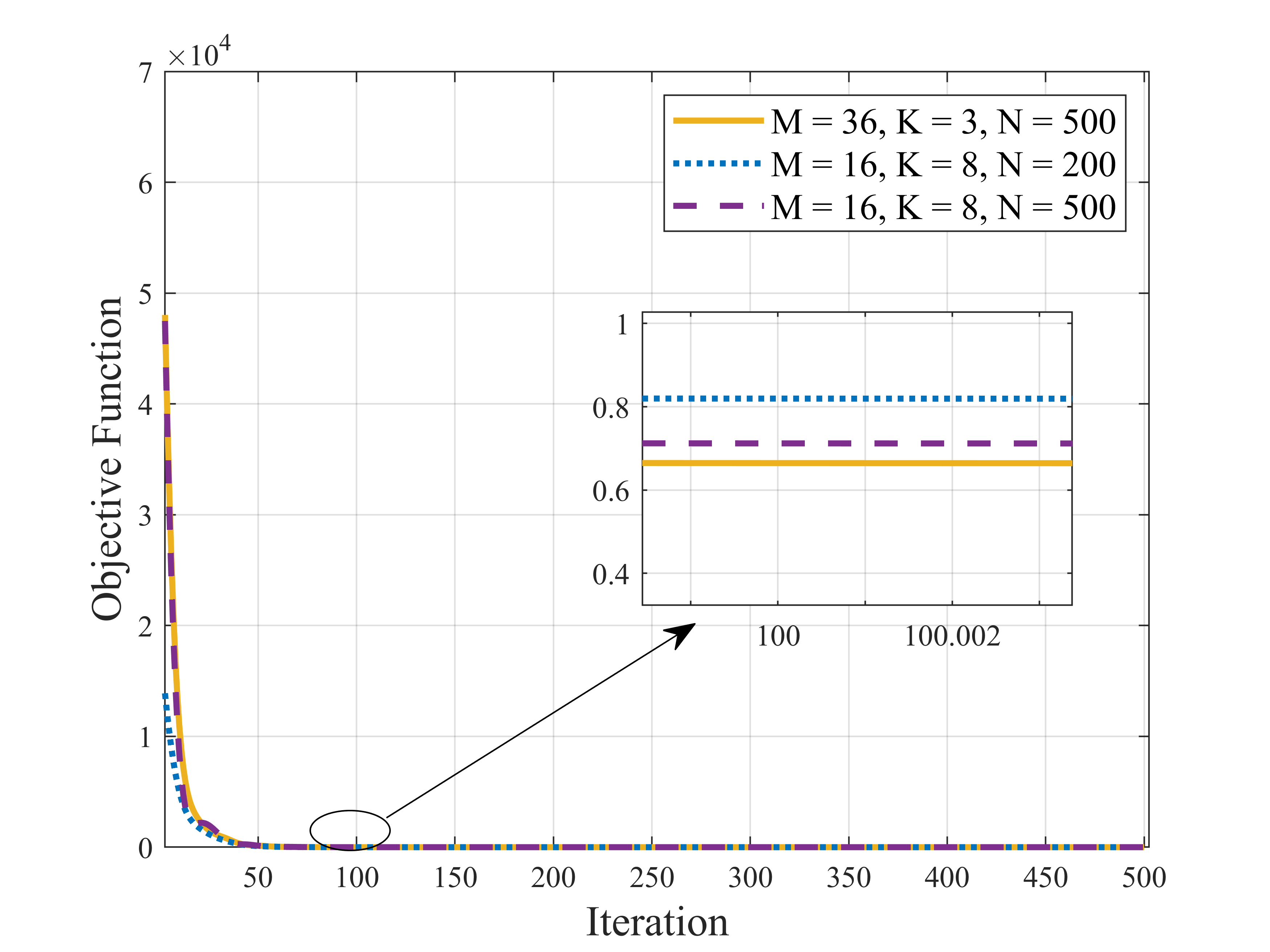}
	\caption{Convergence of objective function.}
	\label{convergence}
\end{figure}

To validate the convergence of our proposed method, we depict the objective function of Problem (\ref{optimization}) in Fig. \ref{convergence}. As can be seen, our proposed method based on Adam can converge to a locally optimal solution rapidly within 100 iterations, which validates the effectiveness of our proposed method.

\subsection{Performance Comparison}
\begin{figure}[t]
    \centering
    \subfigure[4-PAM, \(M \times K = 36 \times 3\)]{\begin{minipage}[b]{0.45\textwidth}
        \centering
        \includegraphics[width=2.5in]{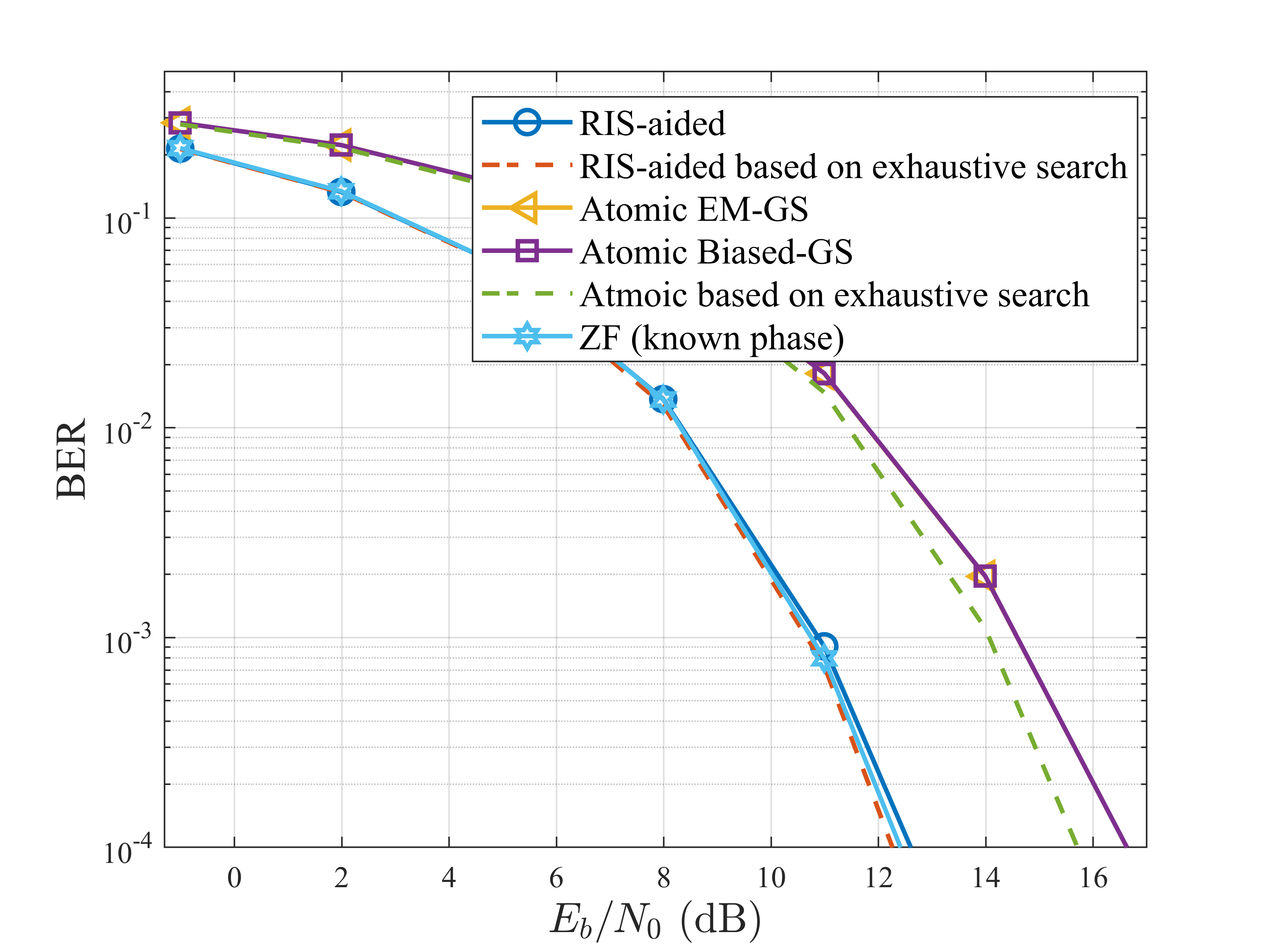}
        \label{4QAMfiga}
    \end{minipage}}

    \subfigure[4-QAM (PAM), \(M \times K = 36 \times 3\)]{ \begin{minipage}[b]{0.45\textwidth}
        \centering
        \includegraphics[width=2.5in]{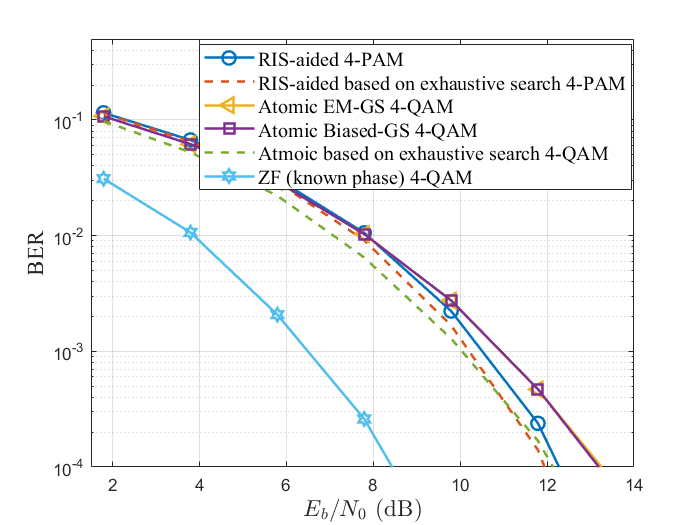}
        \label{4QAMfigb}
    \end{minipage}}
   
    \caption{The effect of \(E_b/N_0\) on the performance with 3 users and 36 vapor cells (a) 4-PAM  and (b) 4-QAM (PAM). Although our proposed atomic receiver has the same performance as the radio frequency one (i.e., ZF with known phase), the atomic MIMO receiver can achieve over 20 dB sensitivity \cite{gong2024rydberg}.}
    \label{4QAM}
    
\vspace{-0.3cm}
\end{figure}

To validate the effectiveness of our proposed detection algorithm, we compare it against the following benchmark methods:

\begin{enumerate}
\item {\bf {Atomic EM-GS}} \cite{cui2025towards}: An iterative signal detection algorithm based on maximum likelihood (ML) detection is implemented in a conventional atomic MIMO receiver.
\item {\bf Atomic Biased-GS} \cite{cui2025towards}: A least-squares (LS)-based iterative algorithm is employed in a conventional atomic MIMO receiver.
\item {\bf Atomic Exhaustive Search}: An exhaustive search-based detection algorithm is applied in a conventional atomic MIMO receiver.
\item {\bf RIS-aided Exhaustive Search}: An exhaustive search-based detection algorithm is applied in the proposed RIS-aided atomic MIMO receiver.
\item {\bf ZF (Known Phase)}: An ideal reference case where the receiver has prior knowledge of the phase of \(\mathbf{z}\) and directly performs zero-forcing (ZF) detection to demodulate the received signal.
\end{enumerate}

As illustrated in Fig. \ref{4QAM}, we investigate the effect of $(E_b/N_0)$ on BER performance with various methods. It can be observed that the BER performance is significantly improved with the increasing power of the transmitted information. More importantly, our proposed method outperforms the two methods in \cite{cui2025towards} when using 4-PAM and 4-QAM, respectively. For 4-PAM, it is worth noting that our proposed method is closer to the exhaustive search method and ideal detection (i.e., ZF with known phase) compared to EM-GS and Biased-GS algorithms. This is because our proposed algorithm can directly demodulate the received signal without alternating iterations on phase and symbols, thereby avoiding error propagation caused by iterative signal detection. Furthermore, compared to a conventional atomic MIMO receiver relying on 4-QAM, our proposed receiver based on 4-PAM can obtain 1 dB performance gain compared to existing benchmarks.

\subsection{Multi-user Case}
\begin{figure*}
	\centering
	\subfigure[4-PAM (QAM).]{
		\begin{minipage}[t]{0.3\linewidth}
			\centering
			\includegraphics[width=2.5in]{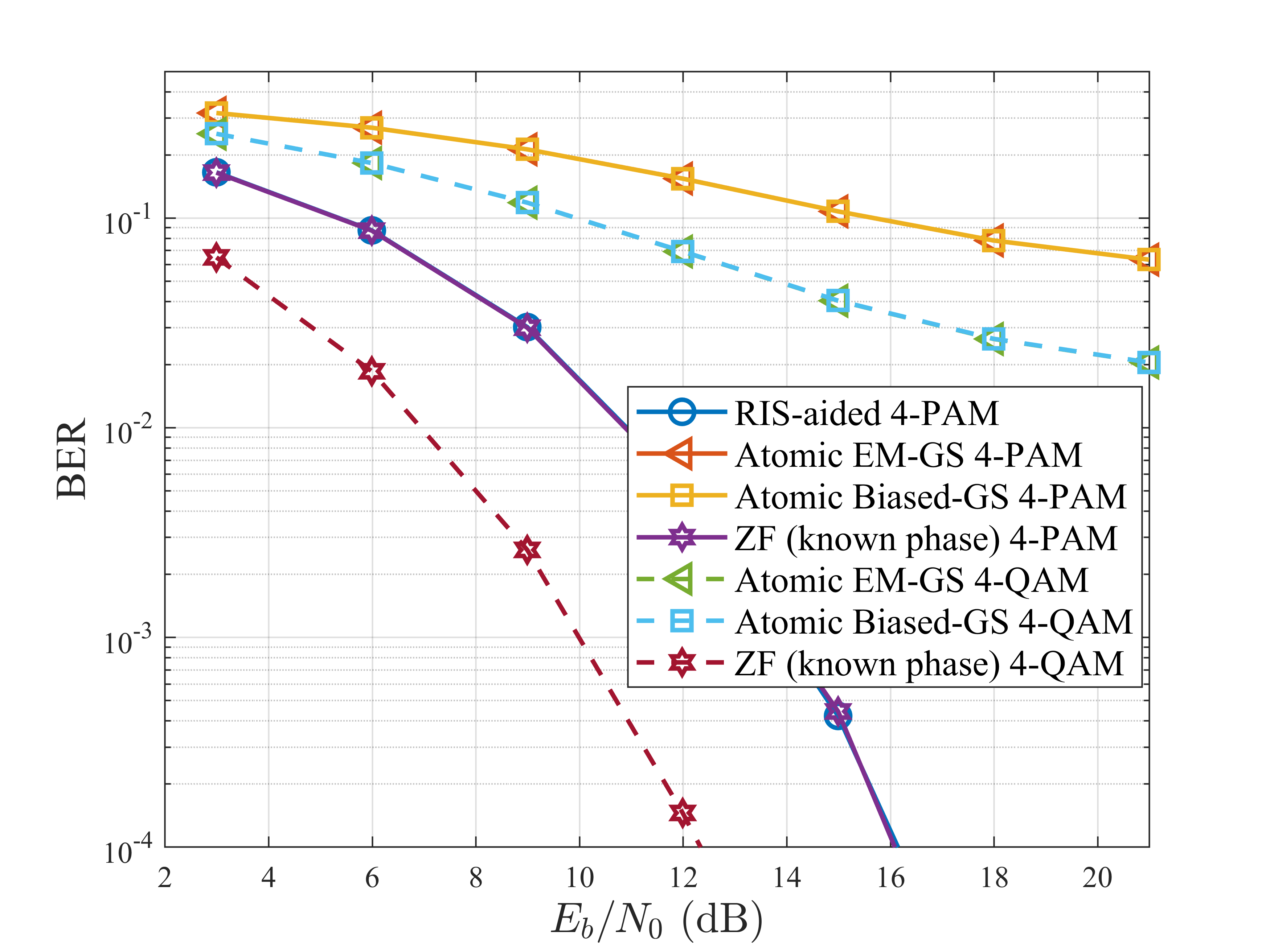}\hspace{10mm}
		\end{minipage}}
	\quad
	\subfigure[8-PAM (PSK).]{\begin{minipage}[t]{0.3\linewidth}
			\centering
			\includegraphics[width=2.5in]{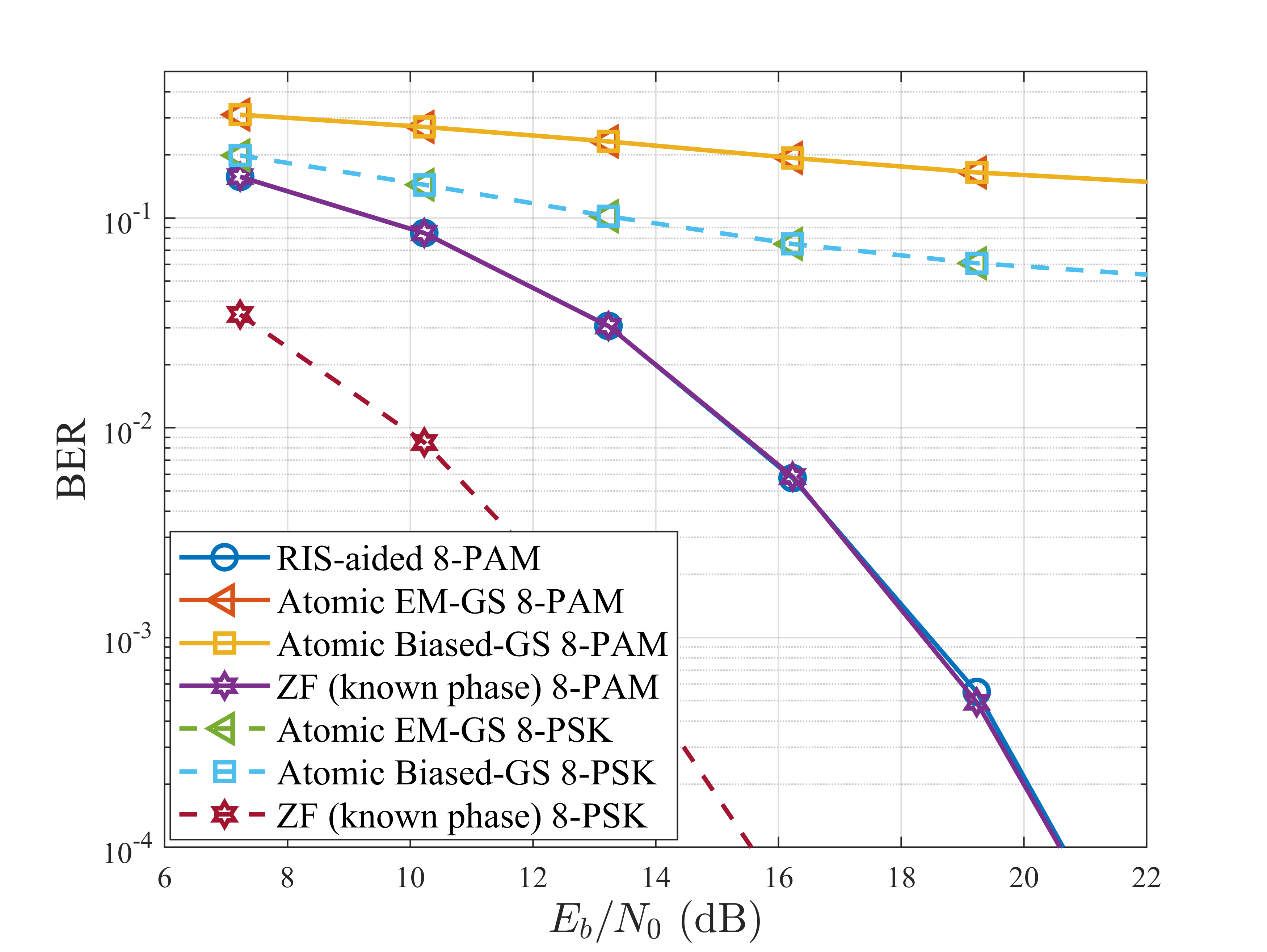}\hspace{10mm}
		\end{minipage}}
	\quad
	\subfigure[16-PAM (QAM).]{\begin{minipage}[t]{0.3\linewidth}
			\centering
			\includegraphics[width=2.5in]{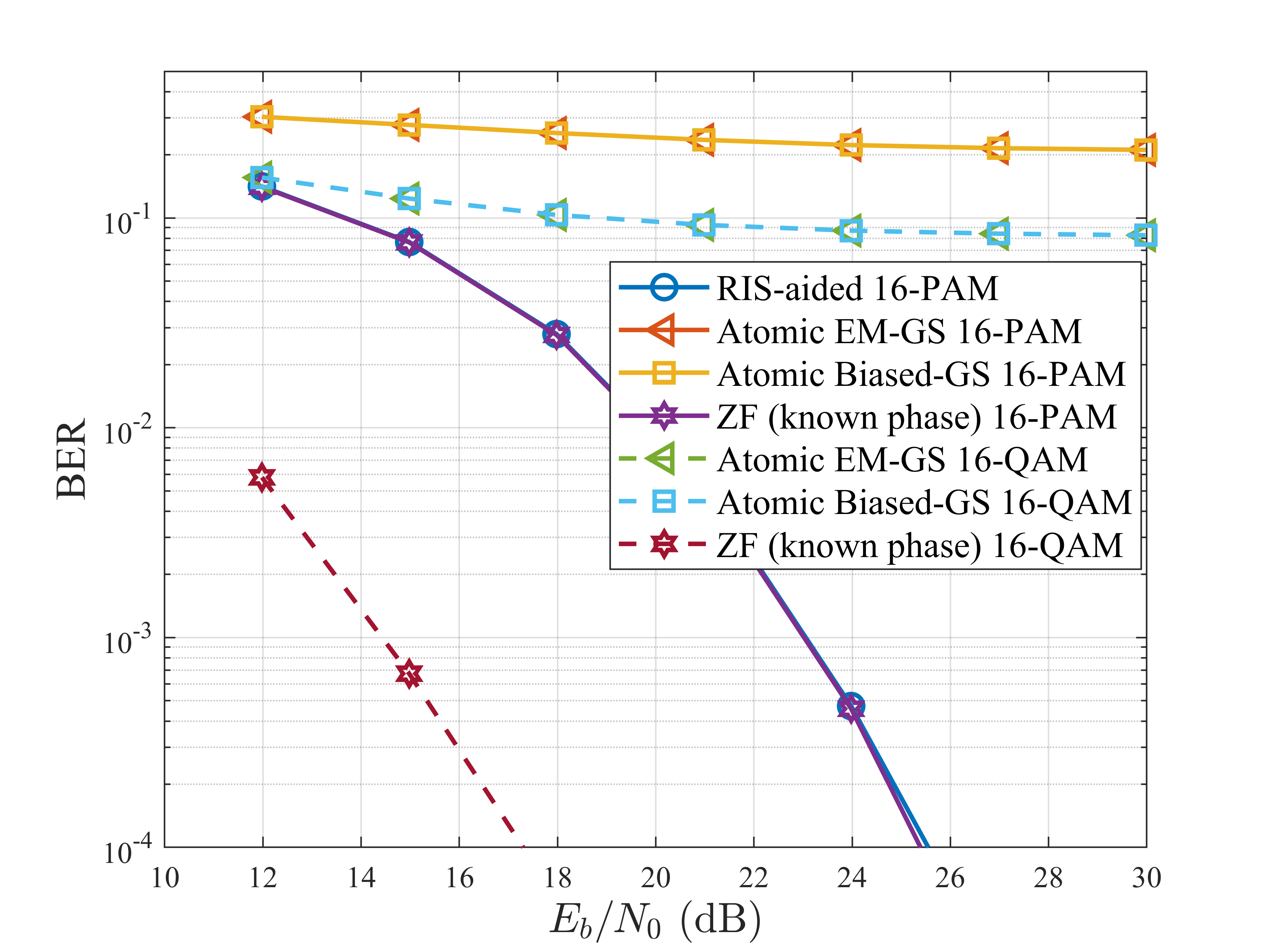} \hspace{10mm}
	\end{minipage}}
	  \caption{The effect of the number of users and vapor cells on the BER performance with \(M \times K = 16 \times 8\)  (a) 4-PAM (QAM), (b) 8-PAM (PSK), and (c) 16-PAM (QAM). Similarly, the atomic MIMO receiver can achieve over 20 dB sensitivity \cite{gong2024rydberg}.}
	\label{cluster}
\end{figure*}

To investigate the impact of the number of vapor cells and users on system performance, we depict the BER performance with 16 vapor cells and 8 users in Fig. \ref{cluster}. Since the complexity of the exhaustive search increases exponentially with the modulation order and the number of users, the performance of signal detection relying on exhaustive search will not be shown. As expected, the ZF method with perfect phase knowledge achieves the best performance, as it can effectively distinguish between multiple users by leveraging the available perfect-phase information, thereby enhancing detection accuracy. Furthermore, we notice that the BERs based on the EM-GS and Biased-GS methods are significantly deteriorated. This limitation arises from the fact that both algorithms rely on highly accurate initial phase estimates to perform iterative searches over symbols and phases. However, in scenarios with a large number of users, accurately estimating the initial phase becomes challenging, resulting in error propagation and accumulation throughout the iterative process. In contrast, our proposed RIS-assisted atomic receiver employs amplitude modulation instead of phase modulation, thereby eliminating phase ambiguity. This approach not only significantly enhances detection performance but also reduces receiver complexity.

\section{Conclusion}
In this paper, we proposed a novel and low-complexity RIS-assisted atomic MIMO receiver. By introducing RIS and employing PAM modulation, the phase of the transmitted signal can be aligned with that of the LO, and then the non-linear signal can be transformed into a linear one, thereby significantly reducing both the signal detection complexity and the overall receiver complexity. To address the non-convex optimization problem, we reformulated it into a tractable form, which can be solved by an Adam-based gradient descent algorithm. Leveraging the optimized equivalent signals, we evaluated the performance of the proposed detection algorithm. The results show that the proposed architecture substantially improves detection performance, achieving a 4 dB gain with 4-PAM in the 3-user case and a remarkable performance improvement of over 10 dB in multi-user scenarios compared to conventional approaches. These gains clearly demonstrate the scalability and effectiveness of the proposed RIS-assisted atomic MIMO receiver in a high-user-density environment.

\begin{appendices}	

\section{Proof of Lemma \ref{lemma1}}
\label{Proof_lemma1}
For minimizing \(\left\|\Im\left\{(\mathbf{A}\mathbf{\Phi}\mathbf{B}+\mathbf{C})\right\}\right\|^2_F\), we notice that the only solution is \(\mathbf{\Phi}^{*}\), satisfying \(\Im\left\{(\mathbf{A}\mathbf{\Phi}\mathbf{B}+\mathbf{C}) \right\} = \mathbf{0} \in \mathbb{R}^{M \times K}\). In the following, we discuss two cases.

Case I: If \(\mathbf{b} \in \mathbb{R}^{M \times 1}\), we have \(e^{j\angle \mathbf{b}} = [{1,\cdots,1}]^T \in \mathbb{R}^{M \times 1}\) and \(\mathbf{s} \in \mathbb{R}^{K \times 1}\). Then, owing to the random information of \(\mathbf{s}\), the only solution for minimizing \( \left\|\Im\left\{(\mathbf{A}\mathbf{\Phi}\mathbf{B}+\mathbf{C})\mathbf{s} \circ e^{-j\angle \mathbf{b}}\right\}\right\|^2_2\) is \(\mathbf{\Phi}^{*}\) as well. Therefore, the two problems are equivalent.

Case II: For \(\mathbf{b} \in \mathbb{C}^{M \times 1}\), we first define \(\mathbf{A}\mathbf{\Phi}^{*}\mathbf{B}+\mathbf{C}\) as \(\mathbf{H}^{\text{Re}} \in \mathbb{R}^{M \times K}\) and the \(m\)-th row of  \(\mathbf{H}^{\text{Re}}\) as \((\mathbf{h}^{\text{Re}}_m)^T \in \mathbb{R}^{1 \times K}\). Then, we can construct a matrice, which is given by
\begin{equation}
    [\mathbf{h}^{\text{Re}}_1\times e^{j\angle b_1}, \cdots, \mathbf{h}^{\text{Re}}_M\times e^{j \angle b_M} ]^T \triangleq \boldsymbol{\chi}.
\end{equation}
Since the transmitted symbols are real values, it is readily proven 
\begin{equation}
    \left\|\Im\{\boldsymbol{\chi} \mathbf{s} \circ e^{-j\angle \mathbf{b}}\}\right\|^2_2 = 0.
\end{equation}
Therefore, by minimizing \(\left\|\Im\left\{(\mathbf{A}\mathbf{\Phi}\mathbf{B}+\mathbf{C})\right\}\right\|^2_F\), we can obtain \(\boldsymbol{\chi}\) and then the corresponging solution \(\mathbf{\Phi}\) for minimizing \( \left\|\Im\left\{(\mathbf{A}\mathbf{\Phi}\mathbf{B}+\mathbf{C})\mathbf{s} \circ e^{-j\angle \mathbf{b}}\right\}\right\|^2_2\) by 
\begin{equation}
    \mathbf{\Phi} = (\mathbf{A}^H\mathbf{A})^{-1}\mathbf{A}^H(\boldsymbol{\chi}-\mathbf{C})\mathbf{B}^H(\mathbf{B}\mathbf{B}^H)^{-1}.
\end{equation}


Based on the above discussions, we complete the proof of Lemma \ref{lemma1}.
\end{appendices}	


\bibliographystyle{IEEEtran}
\bibliography{myref}

\end{document}